%% file: 0-MAIN.tex
\documentclass[conference]{IEEEtran}
\usepackage{cite}
\usepackage{amsmath,amssymb,amsfonts}
\usepackage{algorithmic}
\usepackage{graphicx}
\usepackage{textcomp}
\usepackage{xcolor}
\usepackage{tabularx}
\usepackage{graphicx}
\usepackage{colortbl}
\usepackage{caption}
\usepackage{booktabs}

\usepackage{hyperref}  % Add this in your preamble
\hypersetup{
    colorlinks=true,
    linkcolor=blue,
    filecolor=magenta,      
    urlcolor=blue,
}

\usepackage{fancyhdr}
\def\BibTeX{{\rm B\kern-.05em{\sc i\kern-.025em b}\kern-.08em
    T\kern-.1667em\lower.7ex\hbox{E}\kern-.125emX}}

\begin{document}

\title{ Whole-Person Education for AI Engineers}

\author{
\IEEEauthorblockN{Rubaina Khan}
\IEEEauthorblockA{\textit{University of Toronto} \\
Toronto, Canada \\
rubaina.khan@mail.utoronto.ca}
\and
\IEEEauthorblockN{Tammy Mackenzie}
\IEEEauthorblockA{\textit{Aula Fellowship for AI} \\
Montreal, Canada \\
tammy@theaulafellowship.org}
\and
\IEEEauthorblockN{Sreyoshi Bhaduri}
\IEEEauthorblockA{\textit{Private Corporation} \\
New York, USA \\
sreyoshibhaduri@gmail.com}
\and
\IEEEauthorblockN{Animesh Paul}
\IEEEauthorblockA{\textit{University of Georgia} \\
USA \\
Animesh.Paul@uga.edu}
\and
\IEEEauthorblockN{Branislav Radeljić}
\IEEEauthorblockA{\textit{Aula Fellowship for AI} \\
London, United Kingdom \\
branislav.radeljic@gmail.com}
\and
\IEEEauthorblockN{Joshua Owusu Ansah}
\IEEEauthorblockA{\textit{Arizona State University} \\
USA \\
jowusuan@asu.edu}
\and
\IEEEauthorblockN{Beyza Nur Guler}
\IEEEauthorblockA{\textit{Virginia Tech} \\
USA \\
bng@vt.edu}
\and
\IEEEauthorblockN{Indrani Bhaduri}
\IEEEauthorblockA{\textit{PARAKH, NCERT} \\
India \\
IndraniBhaduri@gmail.com}
\and
\IEEEauthorblockN{Rodney Kimbangu}
\IEEEauthorblockA{\textit{Virginia Tech} \\
USA \\
rodneykb@vt.edu}
\and
\IEEEauthorblockN{Nils Ever Murrugarra Llerena}
\IEEEauthorblockA{\textit{University of Pittsburgh} \\
Pennsylvania, USA \\
NEM177@pitt.edu}
\and
\IEEEauthorblockN{Hayoung Shin}
\IEEEauthorblockA{\textit{Semyung University} \\
South Korea \\
stella0593@gmail.com}
\and
\IEEEauthorblockN{Lilianny Virguez}
\IEEEauthorblockA{\textit{University of Florida} \\
USA \\
lilianny.virguez@ufl.edu}
\and
\IEEEauthorblockN{Rosa Paccotacya-Yanque}
\IEEEauthorblockA{\textit{Universidad Católica San Pablo} \\
Peru \\
rypaccotacya@ucsp.edu.pe}
\and
\IEEEauthorblockN{Thomas Mekhaël}
\IEEEauthorblockA{\textit{École de technologie supérieure} \\
Canada \\
thomas.mekhael@etsmtl.ca}
\and
\IEEEauthorblockN{Allen Munoriyarwa}
\IEEEauthorblockA{\textit{Walter Sisulu University} \\
South Africa \\
amunoriyarwa@wsu.ac.za}
\and
\IEEEauthorblockN{Leslie Salgado}
\IEEEauthorblockA{\textit{University of Calgary} \\
Canada \\
leslie.salgadoarzuag@ucalgary.ca}
\and
\IEEEauthorblockN{Debarati Basu}
\IEEEauthorblockA{\textit{Embry-Riddle Aeronautical University} \\
USA \\
basud@erau.edu}
\and
\IEEEauthorblockN{Curwyn Mapaling}
\IEEEauthorblockA{\textit{Nelson Mandela University} \\
South Africa \\
Curwyn.Mapaling@mandela.ac.za}
\and
\IEEEauthorblockN{Natalie Perez}
\IEEEauthorblockA{\textit{Private Corporation} \\
Hawai‘i, USA \\
natalie.perez@hawaii.edu}
\and
\IEEEauthorblockN{Yves Gaudet}
\IEEEauthorblockA{\textit{Private Corporation} \\
Mirabel, Canada \\
yves@colplastinc.com}
\and
\IEEEauthorblockN{Paula Larrondo}
\IEEEauthorblockA{\textit{Queens University} \\
Canada \\
p.larrondo@queensu.ca}
}

\maketitle

\input{1-Abstract.tex}

\begin{IEEEkeywords}
Whole-person education, Transformative education, Artificial Intelligence, Engineering Education
\end{IEEEkeywords}
\input{2-Introduction.tex}
\input{3-LiteratureReview.tex}
\input{5-Method.tex}

\input{6-Results.tex}
\input{7-Discussion}
\input{8-Conclusion.tex}

% \section*{Acknowledgment}

% %\section*{References}
\bibliographystyle{IEEEtran}
\bibliography{references.bib}
%\begin{thebibliography}{00}
%\end{thebibliography}

\input{PracticeDetails.tex}

\end{document}

%% file: 1-Abstract.tex
%https://ti0ti5.fie-conference.org/authors/full-wip-guidelines

\begin{abstract}

% Should be 300 to 400 words.
This autoethnographic study explores the need for interdisciplinary education spanning both technical and philosophical skills - as such, this study leverages whole-person education as a theoretical approach needed in AI engineering education to address the limitations of current paradigms that prioritize technical expertise over ethical and societal considerations. Drawing on a collaborative autoethnography approach of fourteen diverse stakeholders, the study identifies key motivations driving the call for change, including the need for global perspectives, bridging the gap between academia and industry, integrating ethics and societal impact, and fostering interdisciplinary collaboration. The findings challenge the myths of technological neutrality and technosaviourism, advocating for a future where AI engineers are equipped not only with technical skills but also with the ethical awareness, social responsibility, and interdisciplinary understanding necessary to navigate the complex challenges of AI development. The study provides valuable insights and recommendations for transforming AI engineering education to ensure the responsible development of AI technologies.\\

\end{abstract}

%% file: 2-Introduction.tex
\section{Introduction}
\label{sec:intro}
Traditional technical education puts emphasis on control, utility, and precision. It prioritizes efficiency and optimization through the mastery of tools, algorithms, and design processes \cite{pawley20091b71e, vanasupa20201c103}. Although these hard skills are undeniably valuable, they seem to be taught in isolation from the cultural and ethical operational contexts of these technologies \cite{orr199239682, nussbaum201030239}. Whole-person education, however, is grounded in the development of the learners while integrating technical expertise with interdisciplinary thinking and social responsibility; it is not de-personalized \cite{podger2010c94d3, vanasupa20201c103}. Engineers’s decisions have far-reaching implications for humanity’s well-being; thus, through holistic education, they are not considered only problem-solvers but agents of societal transformation \cite{tronto19934e2bd}. Where traditional models favor disciplinary silos, whole-person education invites pluralism, reflexivity, and a commitment to the common good \cite{crawford2009652b0}. This contrast underscores the need for a transformative shift in AI engineering education, a shift that prepares future engineers not just to build technology, but to ethically navigate it and shape societal impacts by embedding human values into the design process \cite{umbrello2021b002c}. This gap has led to calls for a paradigm shift in engineering education, one that moves beyond a narrow focus on technical proficiency and embraces a more comprehensive approach known as whole-person education. 

Whole-person education emphasizes the holistic development of individuals, integrating technical skills with ethical reasoning, social responsibility, and ecological awareness \cite{chan20228a58b}. In the context of engineering education, this approach recognizes that engineers are not merely builders of technology but also shapers of society, and their work has profound implications for the well-being of individuals, communities, and the planet. Such approaches need educational reform that can support a holistic curriculum \cite{miller2019holistic, edwards2023work}. Whole-person education emphasizes the understanding of practical application of basic science principles \cite{maksiutov2024modern}. It also aims at nurturing the practical application of basic principles of science designed to produce a skilled craftsperson, equipped with technical knowledge \cite{izvorska2022modern}.

Our paper explores the need for a whole-person education approach in AI engineering education, drawing on autoethnographic reflections from diverse stakeholders, including students, educators, and industry professionals. Through their personal and professional experiences, we gain insights into the limitations of current educational paradigms and the urgent need for change. The study challenges the prevailing notions of technological neutrality and technosaviourism, arguing that AI technologies are not value-free tools but rather embody and perpetuate societal biases and power structures. 

Our paper is guided by two primary research questions:
\begin{itemize}
\item RQ1: For whole-person education advocates, what factors influence their support for integration into engineering education, especially as it pertains to AI?\\
\item RQ2: How do the collective reflections of participants envision and shape the future of engineering education, especially as it pertains to AI?
\end{itemize}

The rest of the paper is structured as follows: First, we provide a literature review background on the concepts of technological neutrality and techno-saviourism, highlighting their limitations in the context of AI development. Second, we present the methodology of the study, including the rationale for using collaborative autoethnography and the characteristics of the author-participants. Third, we analyze the key themes emerging from the author-participants' reflections, exploring their motivations for advocating for a whole-person education approach and their vision for the future of AI engineering education. Finally, we discuss the implications of the findings and provide recommendations for integrating whole person education principles into engineering programs.

%% file: 3-LiteratureReview.tex
\section{\large{Literature Review}}
\label{sec:related}
\subsection{Technological Neutrality and Technosaviourism}

The ethical and societal implications of artificial intelligence (AI) are shaped by two pervasive but contested ideologies: technological neutrality and technosaviourism. Technological neutrality is the principle that regulation should not impose different rules based on the technical means used to provide a service but should instead focus on the service's function and its societal impact. Neutrality is defined as “the freedom of individuals and organizations to choose the most appropriate technology adequate to their needs and requirements for development, acquisition, use or commercialization, without knowledge dependencies involved as information or data” \cite{rios2013technological}. The assumption with neutrality is that technology can be neutral because it is designed to be so. However, in actual practice, it is difficult to maintain neutrality. Often the programmers who design neutral technology are hidden behind their algorithms. These frameworks, rooted in the philosophy of technology, influence how engineers conceptualize AI systems and their societal role.

The claim that technologies are neutral, mere tools whose impacts depend solely on human use, has long been disputed in philosophy of technology.Seminal critique argued that technologies inherently embody political values, as seen in designs that enforce social hierarchies (e.g., nuclear power infrastructure) \cite{winner1978autonomous, winner2017artifacts, buolamwini2018c3039}. Contemporary scholars extend this critique to AI, emphasizing how algorithmic systems encode biases through training data, design choices, and institutional priorities \cite{benjamin2019ce394, birhane202188b83}. Similarly, Crawford and Joler \cite{crawford201848a5b} deconstruct AI’s “myth of neutrality” by tracing its ecological and labor costs, such as energy-intensive data centers and exploitative mining practices for hardware materials. These critiques underscore how AI systems are socio-technical artifacts, inseparable from the power structures that produce them \cite{zuberi2010critical}.

Technosaviourism is the belief that technologies can solve major social problems. In the field of AI, it is an ideology that has been documented by scholars and is espoused by many of the people who have the most control over the technologies and companies that control AI \cite{gebru2024tescreal}. Often, technosaviourism is also coupled with a belief that the controllers of technology know what is best for all of society, and that they should be the ones to enact major changes, not governments.

\subsection{Ethical Frameworks for Engineers Working on AI}
The lack of transparency is at the core of AI  harm. The rigidity of training data, model architectures, and decision-making logic often shields discriminatory patterns from scrutiny and auditing, paving the way for biases to remain unchallenged \cite{pasquale2015f9193, diakopoulos2016a38c7}. This impermeable nature of many AI systems contradicts the concept of technological neutrality and instead reveals the dominant processes that shape algorithmic outcomes \cite{ananny201889bec}. Transparency, therefore, is not only a technical feature. It is a prerequisite for public trust \cite{selbst2018d7665}. 

In many engineering programs, however, transparency is scarcely framed as an issue of code documentation rather than a moral or civic concern. A whole-person education approach would reconceptualize transparency as a design principle. One that empowers users and engineers alike to expose biases and democratize knowledge. Frameworks such as Value Sensitive Design \cite{umbrello2021b002c} and Inclusive Design \cite{clarkson2013inclusive} advocate for integrating transparency at the foundational stages of system development. Thus, embedding transparency into AI engineering education can address the gap in current engineering curricula and transform it into a fundamental ethical competency. As AI continues to shape society and influence critical decisions affecting people’s lives, engineering education must integrate frameworks to ensure ethical development of AI. The field of Human-Computer Interaction tackles questions regarding responsible and fair AI development. However, curriculums usually lack such frameworks. 

This section provides an overview of three key frameworks - inclusive design, participatory design and value sensitive design. First of all, inclusive design in HCI has ethical and social roots. It is related to universal design where everyone has the right to access the same services. Inclusive design incorporates the user in the design process not just at the final product, but from the initial conception of the design idea \cite{abascal2007517fe}. 

In addition, the Canadian inclusive design framework challenges typical design processes that view accessibility as a mere checklist. In addition, the founder of the Inclusive Design Research Centre (IDRC) and framework, Jutta Treviranus \cite{treviranus2019value}, extends on this by arguing that decision making should not be based on principles behind “Big Data” such as centering most of the customer’s needs (80\%). This center approaches inclusive design through three dimensions which acknowledges and respects human variability, promotes the inclusion of diverse perspectives especially for those who can’t or have difficulty utilizing a specific design, and prioritizes decisions that benefit all. The practitioners of the inclusive design framework are seeking to find models that can continuously personalize systems according to the user’s needs instead of a one-size-fits-all approach. 

Secondly, the participatory design framework has democratic roots for settings beyond the workplace \cite{bødker2004e9626}. Participatory design essentially puts the users who are impacted by a specific technology or product on the forefront of design as co-designers\cite{treviranus2019value}. In addition, participatory design promotes empowerment through design processes \cite{bødker2004e9626}). More specifically, bridging between technology developers and users results in mutual learning that combines diverse perspectives and knowledge. This information transfer between the user and technology world also helps with conveying alternative solutions instead of one mainstream solution which mediates the conversation on what is technologically feasible \cite{bødker2004e9626, muller2012participatory}. As a result, users are empowered with understanding the knowledge and problems that come with technology which is facilitated through the recognition of users as resourceful and skillful. 

Lastly, value-sensitive design (VSD) is a systematic approach to incorporating human values in the design process \cite{friedman2017survey}. This framework provides a holistic approach towards design by bringing technical and socio-structural design spaces together. A key strength of VSD is its broader perspective towards design by identifying and justifying the direct and indirect stakeholders to identify who is the most affected by design decisions \cite{friedman2017survey,trewin2019considerations}. Additionally, VSD seeks to resolve value tensions constructively among stakeholders and encourages designers to continuously improve systems. Overall, VSD promotes human well-being in design systems by acknowledging human values as a design criterion.

\subsection{Engineers Building a Relationship with AI - Intellectual Interdependence}

AI prompts debates about the limits of human exceptionalism \cite{boyle2024f9288}. When GPT is approached with questions about its potential to serve, if not replace, a public intellectual, it acknowledges its readiness to engage with complex ideas despite emphasizing its limitations in terms of original thought or advocacy for social change \cite{radeljicai, mackenzie2025we}. Moreover, it goes on to display preference for neutrality over taking stances on issues, which raises concerns. AI’s seeming neutrality could contribute to inequality, especially in the context of the divide between the Global North and the Global South \cite{kj2024decoding}, since AI is accessible only to those with reliable internet access, leaving one-third of the global population without such access. This disparity highlights that AI’s seeming neutrality may obfuscate power structures, limiting access to knowledge and reinforcing exclusion. Additionally, the increasing influence of AI in shaping political discourse and decision-making further underscores its potential to prioritize certain agendas over others. AI models can provide accurate position estimates that influence political actors, potentially shaping the public sphere in ways that serve specific interests, whether they be governmental, corporate, or ideological. 

\subsection{Towards Critical Engineering Education}
The critiques of neutrality and technosaviourism reveal urgent gaps in AI education. Current technical curricula often neglect the sociotechnical dimensions of AI, treating ethics as an add-on rather than a core design consideration \cite{veliz2021moral}. To counter this, scholars advocate for pedagogical frameworks that integrate critical theory, ethics, and interdisciplinary collaboration. For example, value-sensitive design methodologies encourage engineers to proactively identify how AI systems might amplify inequities or environmental harm \cite{umbrello2021b002c}. Case studies of AI failures such as racial bias in facial recognition \cite{buolamwini2018c3039} or the ecological costs of cloud computing \cite{ligozat2022unraveling} can bridge technical and ethical reasoning. Additionally, fostering partnerships with social scientists and communities impacted by AI helps engineers confront the limitations of technosaviourism \cite{schiff2020principles}, and such collaborations shift engineering education from “solutionism” to stewardship, emphasizing accountability over technical prowess alone .

\subsection{Whole-Person Education Models}
Transparency plays a crucial role in cultivating ethical stewardship among future AI engineers. As AI systems increasingly shape decisions in high-stakes domains, including finance and law enforcement, engineers must be prepared to develop models and account for how the models function, their limitations, and who they may affect. This goes beyond technical explainability; it requires a critical awareness of bias in training data, the social implications of model outputs, and the uneven distribution of AI’s harms and benefits.Frameworks like Value Sensitive Design (VSD) and Inclusive Design emphasize transparency not merely as a desirable trait but as a structural design principle that supports stakeholder accountability, contextual responsiveness, and the active surfacing of value conflicts \cite{umbrello2021b002c, treviranus2019value}. Embedding these frameworks into engineering curricula allows transparency to move from a compliance-oriented mindset to a transformative ethical capacity. Additionally, teaching transparency in parallel with power-aware design encourages engineers to see themselves as neutral builders and as socially accountable stewards of sociotechnical systems \cite{ananny201889bec, selbst2018d7665}. Within a whole-person education approach, transparency becomes essential for empowering students to ethically navigate complexity and design AI systems that are both technically sound and socially just. For instance, a student designing an AI diagnostic tool should be taught to document the model’s performance metrics, its potential failure modes, and the demographic limitations of its training data, practices emphasized in both VSD and Inclusive Design pedagogy.

Podger, Mustakova-Possardt, and Reid \cite{podger2010c94d3} advocate for a whole-person approach to educating for sustainability (EfS), arguing for the cultivation of moral motivation and higher-order dispositions, which they see as essential for developing a globally responsible consciousness.  This approach is presented as an alternative to the prevalent focus on specific capabilities and competencies in higher education.  The authors draw on psychological research to emphasize the importance of integrating the development of the powers of mind, heart, and will, fostering a consistency between what learners know, what they care about, and the choices they make.  They highlight that this form of education makes possible the fulfillment of one's human and spiritual potential, and is essential for the individual and cultural change towards sustainability. Others discuss the awakening of engineering education in response to the challenges posed by the COVID-19 pandemic \cite{vanasupa20201c103}, emphasizing the urgent need to prioritize public health and well-being over economic benefits. Authors critique the conventional engineering education model, which underplays emotions and overlooks the mental health of students, suggesting that this approach contributes to widespread mental illness among college students and working professionals. Furthermore, many advocate for the creation of curricula that honor the dignity of all individuals, integrating emotional intelligence, empathy, and social responsibility into engineering education. They call for a collaborative redesign of engineering curricula, encouraging educators to listen to marginalized voices and embrace whole learning practices that foster a sense of community and inclusivity \cite{mackenzie2024reimagining, mackenzie2025we}. Ultimately, they argue for a transformation in engineering education that aligns with societal needs, aiming to cultivate a future where all individuals can thrive and contribute positively to the planet. This kind of transformation is especially relevant for engineers working on AI - where whole-person models are critical for developing ethical discernment in automated decision-making, fostering empathy in human-AI interactions, and guiding responsible design.

In the discourse on engineering education, there is a growing recognition of sustainable development not merely as an 'aspect,' but as a pivotal meta-context for curriculum renewal. There have been discussions on the need for a shift towards viewing sustainable development through a holistic, system-wide lens within engineering education \cite{mulder2013sustainable} to equip both students and educators with the capacity to navigate complex dilemmas arising from conflicting sustainability needs in design and research. By advocating for sustainable development as a meta-context, this research underscores the significance of strategic capacity building in engineering education to address the challenges posed by sustainability considerations. In AI education, this systemic view is crucial for grappling with the ethical, environmental, and social trade-offs inherent in the development and deployment of intelligent systems. For example, in their vision for interdisciplinary engineering education, Van den Beemt \cite{van2020interdisciplinary} observed that one of the important roles for interdisciplinary education efforts is to help students develop the kind of flexible, adaptive expertise that will prepare them to solve a range of complex problems and work with scientists trained from a variety of perspectives, as is increasingly becoming the case in cutting-edge research fields.

%% file: 5-Method.tex
\section{Method}

\begin{figure*}
    \centering
    {
        \includegraphics[width=\textwidth]{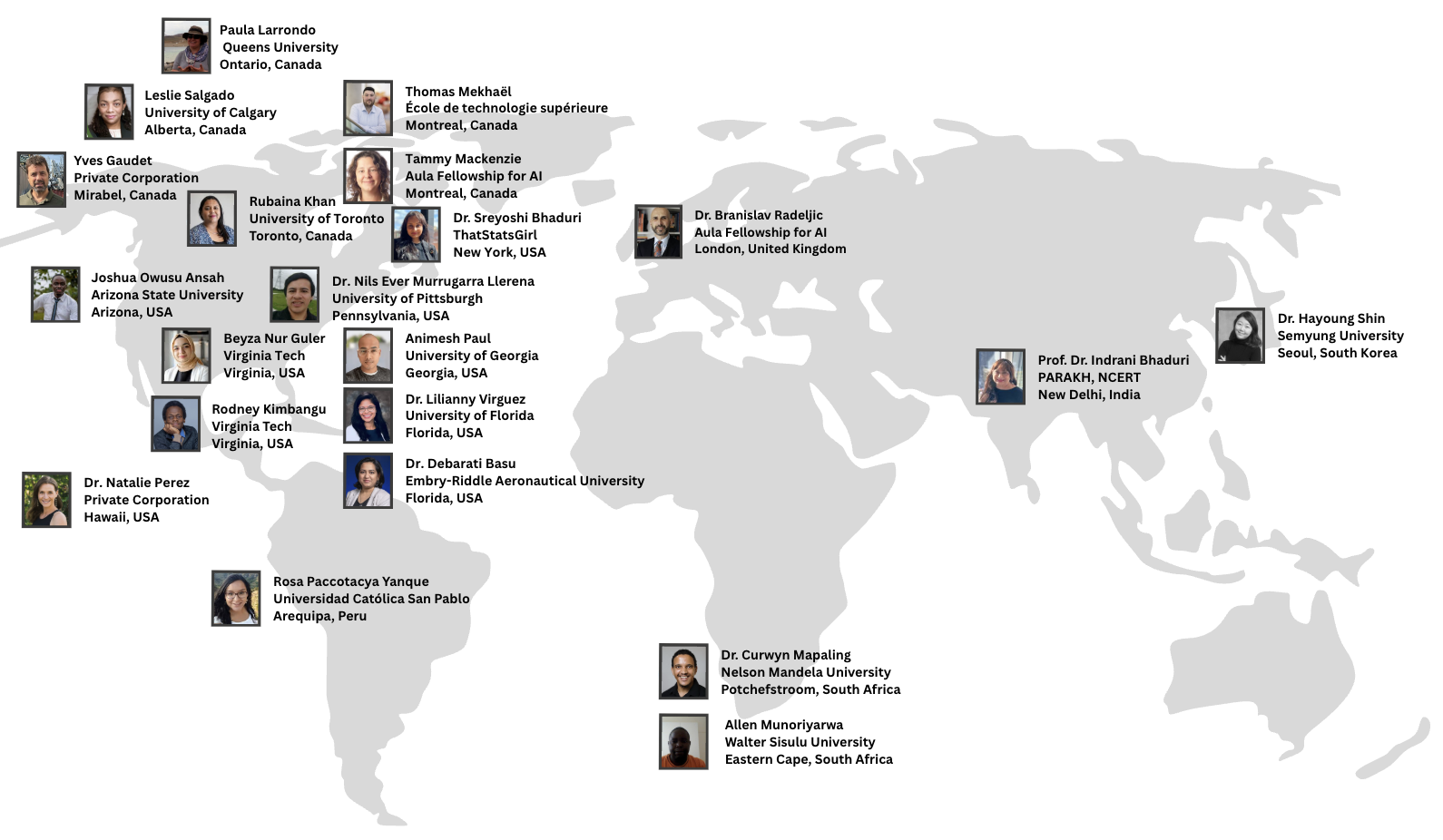}
    }
    \caption{Map visualizing author participants to their current regions of residence}
    \label{fig:enter-label}
\end{figure*}

This study adopted an analytical collaborative autoethnography (CAE) approach, rooted in critical qualitative inquiry, to explore how stakeholders across academia and industry envision a more ethical and holistic AI engineering education. Autoethnography (AE) is a qualitative method that focuses on self as a study subject but transcends a mere narration of personal history \cite{chang2008d0645, chang20165f4d1, battel2021we}. Creswell and Creswell \cite{creswell201747b5e} describe autoethnography as a research methodology that analyzes a phenomenon through the use of self-narratives, which would otherwise remain "private or buried." Autoethnography was an appropriate strategy to use in this study because it provides the participants the opportunity to shift from being outsiders to insiders in the research, which further enables their voices to be better heard within the community, thus promoting convergence and inclusion. 

CAE extends this further by engaging multiple researchers who collectively explore their autobiographical narratives to interrogate broader sociocultural phenomena \cite{chang2008d0645}. As Ellis and Bochner \cite{ellis2000a458b} emphasize, auto-ethnographers vary in their emphasis across the axes of auto (self), ethno (culture), and graphy (research process), and this study situates itself centrally across all three. Like a musical ensemble where each instrument offers a distinct voice, CAE enables participants to contribute unique reflections while building a shared narrative with deeper resonance and criticality. 

The research team consisted of twenty one participants—each with distinct global backgrounds in engineering, education, AI research, and industry practice. This diverse group of participants brings a rich variety of perspectives to the study, ranging from academic to industry viewpoints, and representing different stages of career development in the field of AI and engineering education. Positionality statements were included in our reflective process and informed the thematic interpretation of the narratives. The researchers, as both instruments and data sources, produced written reflections each (approximately 300 average number of words per entry). The reflective prompts elaborated in Appendix  guided the data collection process. 

To accommodate the diverse time zones, schedules, and communication preferences of an international group, twenty participants initially responded asynchronously to research prompts via a shared digital document. This asynchronous method was deliberately chosen to allow individuals ample time for deep reflection on their experiences before crafting their responses, which fostered more thoughtful and authentic contributions. Additionally, the shared document format enabled dialogic engagement; participants could read and respond to one another's narratives over time, thereby enriching the collective reflexivity that is central to autoethnographic inquiry.

Following the collection of these initial reflections, the research team met for two synchronous video conference sessions. The purpose of these meetings was to share interpretations, question underlying assumptions, and deepen the collective analysis of the data. Subsequently, the inquiry was extended by inviting a new set of author-participants to join the group and contribute their thoughts on the topic. These additional reflections augmented the dataset, bringing the total number of contributing participants to twenty.

For data analysis, the team employed thematic analysis, following the six-phase process \cite{braun200612699}: familiarization, coding, generating themes, reviewing themes, defining and naming themes, and writing up the findings. This process was iterative rather than strictly linear, allowing the group to revisit earlier stages as new patterns or insights emerged. This commitment to collective reflexivity fostered deeper learning about both self and other, while simultaneously enhancing the trustworthiness and richness of the findings through practices of power-sharing and community building \cite{chang20165f4d1}.

It is important to note that Collaborative Autoethnography (CAE), the methodology used here, does not aim for generalizability. Instead, it prioritizes insightful depth and relational meaning-making rooted in real-world experiences. This research adopts an idiographic approach, which focuses on understanding individual cases in depth to emphasize the uniqueness within each participant's experiences. This contrasts with a nomothetic approach, common in many scientific fields, which seeks to identify generalizable trends across large populations.

Contrary to framing positionality as a limitation, we affirm that subjectivity is a feature, not a flaw, in our autoethnographic research. Future iterations of this study will include more detailed positionality statements in-text to further contextualize interpretation and reflexivity. At the time of the study, all participants were adults engaged in professional or academic roles and contributed voluntarily to this project. As such, this autoethnographic inquiry did not involve external human subjects, thus institutional ethics review was not applicable under internal authorship; and all participants gave informed consent for inclusion, authorship, and dissemination.

%% file: 6-Results.tex
\section{Results and Discussion}

\begin{figure*}
    \centering       \includegraphics[width=\textwidth,height=0.8\textheight,keepaspectratio]{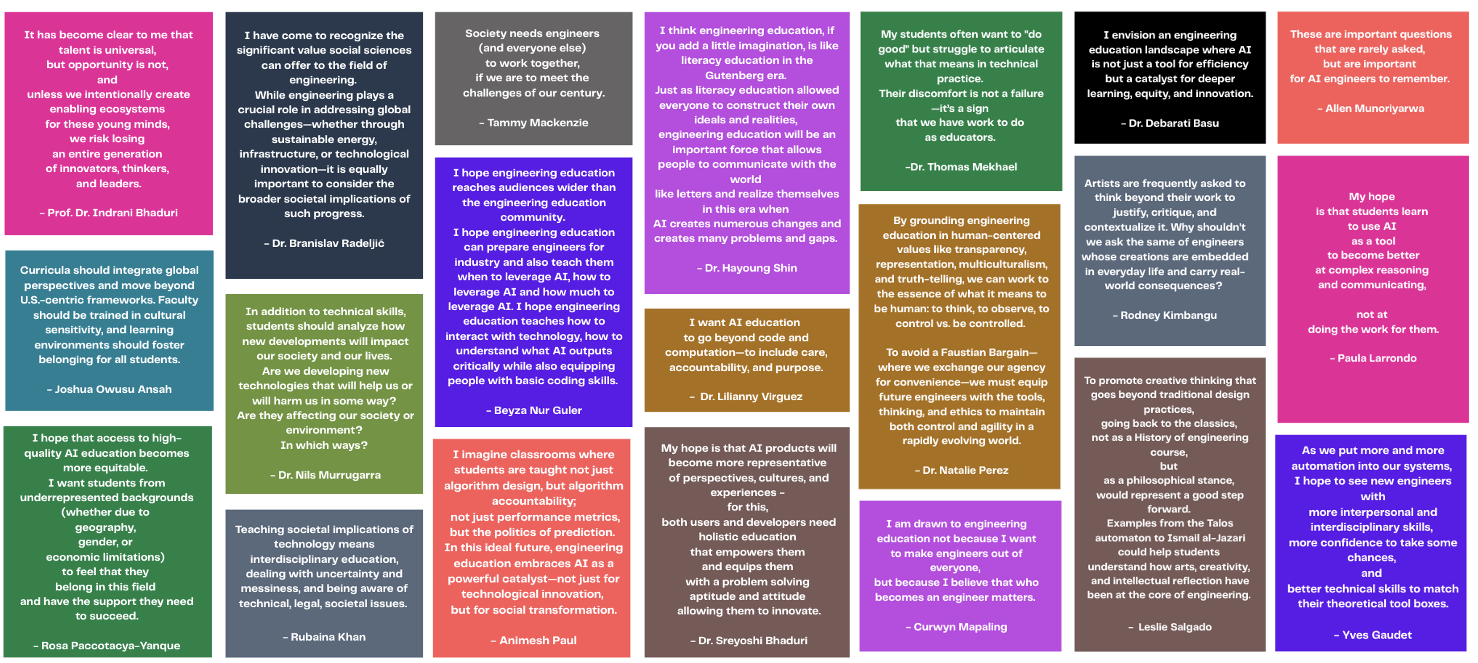}
    \caption{Select snippets of reflections from author-participants, advocating for whole-person education.}
    \label{fig:enter-label}
\end{figure*}

This research explores two interrelated questions: First, what motivates author-participants to advocate for a whole-person education approach in the context of Artificial Intelligence (AI)? Second, how do their collective reflections envision the future of AI engineering education? Drawing on rich, autoethnographic narratives, this section presents the results of an analysis of participant experiences and values, highlighting both individual perspectives and an emerging collective vision. Through a global lens, our findings strongly reaffirm a pressing need for a holistic, interdisciplinary AI education.

\subsection{Key motivations driving participants to advocate for a whole-person education approach for engineers developing AI solutions.}
Across the reflections, participants shared personal and professional experiences that emphasized the urgency of re-imagining AI training in engineering education through a broader, human-centered lens. These motivations crystallized into five interrelated themes:

\subsubsection{Global and Culturally Responsive Education}
Participants articulated a strong desire for AI education that reflects diverse cultural experiences and serves global populations equitably. Joshua’s experience as an African student navigating Western-centric systems underscored how engineering curricula often overlook racial and geographic inclusivity. Nur, raised between Turkish and American educational systems, similarly emphasized the importance of reducing barriers for special needs populations and students from underrepresented backgrounds. Indrani’s extensive work in India with large-scale assessments illuminated how opportunity gaps disproportionately affect girls and rural learners, echoing her call to “reimagine engineering education content and tools to reflect learners’ lived realities.” Branislav, from a political science background, further reinforced the value of integrating global ethics and equity considerations, particularly in addressing extractive industries like lithium mining, where environmental degradation and exploitative labor practices raise pressing ethical concerns. Collectively, these diverse perspectives highlight that addressing learner needs, ensuring equitable access, and embedding global ethics are not peripheral but foundational to meaningful and inclusive education.

\subsubsection{Bridging Academia, Industry, and Everyday Realities}
There was a consistent critique of the disconnect between academic curricula and real-world AI applications. Tammy, a tech CEO, voiced frustrations with engineers often lacking teamwork, communication, and ethical reasoning skills, especially in team situations. Yves had similar concerns. Nur and Rosa both pointed to how their educational journeys revealed a mismatch between classroom preparation and industry expectations, especially regarding practical skills and inclusive design. These reflections aligned with Sreyoshi’s call for AI education to foster “dialogue and cooperation across domains", bridging educators, practitioners, and users in shared problem-solving.

\subsubsection{Ethics and Social Responsibility as Foundational, Not Optional}
Ethics emerged as a core motivation. Rubaina reflected on the invisibility of societal implications in current curricula, emphasizing that “ethical becoming needs time.” Joshua’s story of his mother’s passing being linked to a biased medical device, served as a powerful reminder of how engineering failures can disproportionately harm marginalized communities. Lily recalled a student who failed to consider who might be excluded by their tech solution, prompting her to stress that education must equip engineers to build “not just smart solutions, but just ones.” Animesh and Rosa, too, emphasized AI's potential for community-driven problem-solving and highlighted the ethical duty to build tools that reflect human diversity.

\subsubsection{Interdisciplinary Learning and Humility}
Participants shared a commitment to breaking disciplinary silos. Branislav and Lily noted the need for AI engineers to engage with social sciences and humanities to navigate complexity and ethical nuance. As Lily observed, “technologies are never neutral: they carry assumptions, priorities, and consequences.” Animesh and Sreyoshi called for AI education that teaches students not only how to code, but how to care. Hayoung envisioned engineering education as a canvas where students can create freely empowered by interdisciplinary tools and inspired by AI’s potential to democratize knowledge. Yves, as a manager of engineers, finds it hard to integrate engineers into teams with other technically skilled workers, noting a tendency to split ideas and participation along social class lines, regardless of actual skill levels. 

\subsubsection{Democratization of AI education for all}
Nearly all participants emphasized the urgency of making AI education more accessible and inclusive. Indrani and Rosa highlighted the need to identify and nurture talent from underserved communities through tools like AI-based assessments and foundational STEM supports. Joshua and Sreyoshi called for culturally sensitive, globally relevant curricula. Nur and Lily stressed the importance of workforce readiness for all including those with disabilities and underrepresented identities, thereby linking this to broader democratization efforts in tech development.

\subsection{Collective reflections of participants envisioned a future for engineering education in the era of AI.}

Participants' visions for the future of AI education coalesce around transformative values: equity, accountability, interdisciplinarity, and user-centered innovation. Their reflections paint a future where AI is not merely taught as a technical discipline, but as a deeply human one.

\subsubsection{Challenging Technological Neutrality}

Many participants rejected the idea that technology is inherently neutral. For example, reflections from Animesh, Joshua, and Lily highlighted how AI systems, without critical examination, can perpetuate and even worsen existing inequalities. Expanding on this, Rubaina and Branislav emphasized the crucial need to situate AI within its specific social, political, and environmental contexts. They viewed ethical reasoning concerning AI as both a technical requirement and a moral obligation. These insights collectively suggest that if engineering education neglects to teach learners how to identify the broader social, political, and cultural consequences of AI, we risk exacerbating the very inequalities that these powerful systems can introduce or amplify.

\subsubsection{Moving Beyond Technosaviourism}
Participants rejected the idea of AI as a silver bullet. Tammy’s critique of engineers lacking civic skills point to a growing disillusionment with exclusively technical skill sets. Instead, participants envision AI engineers as civic actors who have the potential to shape technologies that reflect shared values and collective responsibility. Yves for his part hopes to see young engineers have a more realistic understanding of how humans and machines interact in industrial situations where both are required. 

\subsubsection{Centering Interdisciplinarity as a Norm, Not an Exception}
The future of AI education, participants suggest, must dissolve boundaries and better belonging. Sreyoshi, Rosa, and Brani all call for education that synthesizes insights from education, ethics, sociology, and political science. Hayoung likens engineering education to a creative process where imagination, justice, and diversity intersect with algorithms and design.

\subsubsection{Designing User-Centered, Inclusive, and Flexible Curricula}
Participants argued that curricula need to prioritize users, be highly adaptable, and better prepare students for a dynamic world. Highlighting specific aspects, Rosa and Lily emphasized the integration of design justice and accessible technology principles. Complementing this, Joshua, Nur, and Indrani advocated for curriculum reforms that genuinely reflect learners’ diverse contexts and actively foster critical thinking – not just for students, but for educators as well. Lilianny powerfully summarized a desired shift beyond purely technical skills, stating, “I want AI education to go beyond code and computation, to include care, accountability, and purpose.” Consequently, engineering education should incorporate more real-world, problem-based activities. These learning experiences need objectives focused on developing critical thinking and adaptability, while also instilling the practice of designing for others with a strong sense of care, accountability, accessibility, and justice.

\subsubsection{Prioritizing Ethical Leadership and Lifelong Learning}
Beyond technical skill, participants stressed that engineering education must aim to cultivate ethical leaders. Rubaina emphasized the journey of “ethical becoming,” noting that this development requires dedicated time, deep reflection, and supportive mentorship. Underscoring the societal stakes, Hayoung drew a parallel between AI education and the dawn of literacy in the Gutenberg era; she described it as a potentially transformational force that could either empower or exclude, depending critically on the educational approach taken. This perspective highlights that ethics in engineering education cannot be incidental; it must be pursued as a purposeful outcome. The goal should be to empower learners to build solutions relevant to global, social, and local contexts, while also developing them into critical thinkers and leaders engaged in designing holistically for people. All agreed that the future demands engineers who are not only technically skilled, but socially and locally conscious, globally collaborative, and critically aware.

%% file: 7-Discussion.tex
\section{Implications}

This critical auto-ethnographic study explored the motivations driving a global call for a Whole-Person Education (whole-person education) approach within AI engineering education, examining how diverse stakeholders envision its future. Our collective reflections advocate for a fundamental shift away from purely technical training towards a more holistic, ethically grounded, and socially responsible paradigm. This aligns directly with whole-person education frameworks that emphasize education's role in fostering well-rounded individuals equipped for complex challenges \cite{podger2010c94d3, vanasupa20201c103, bhaduri2024multi, carrico2023preparing}. The insights gathered not only challenge prevailing notions of technological neutrality and technosavvy but also demonstrate the critical applicability of whole-person education principles to the unique demands of educating future AI engineers.

A primary theme emerging from participant reflections is the need to move beyond mere technical competence to cultivate ethical leadership and holistic understanding. The strong consensus on integrating ethics, critical theory, emotional intelligence, empathy, and social responsibility resonates deeply with whole-person education models. This call echoes the work of many others who advocate for cultivating moral motivation and integrating the "powers of mind, heart, and will" \cite{podger2010c94d3} to foster consistency between knowing, caring, and acting \cite{khan2024path}. It also aligns with authors'  critique of conventional engineering education's tendency to underplay emotions and calls for curricula that honor individual dignity by integrating emotional intelligence, empathy, and social responsibility \cite{vanasupa20201c103}. Participants' emphasis on ethics training that transcends mere compliance, demanding integrity and transparency, underscores the need for "ethical discernment" and "responsible design," as highlighted by Vanasupa \cite{vanasupa20201c103} in the context of AI, and reflects the "moral motivation" central to the whole-person education framework described by Podger et al. \cite{podger2010c94d3}.

Furthermore, participants stressed the imperative to situate AI development within broader contexts, directly challenging technological neutrality. This resonates with Mulder and colleagues' \cite{mulder2013sustainable} argument for sustainable development—and by extension, broader societal contexts—as a "pivotal meta-context" for engineering curriculum renewal, requiring a "holistic, system-wide lens." The need to address global perspectives, equity, inclusion, and specific contextual challenges, such as the disparities and colonial legacies impacting the Global South \cite{heenkenda2022role} and algorithmic biases \cite{hawkins2018beijing}, aligns with this systemic view. It also connects to calls to listen to marginalized voices and foster inclusivity, and emphasis on developing a "globally responsible consciousness. \cite{podger2010c94d3}" The call for interdisciplinary collaboration further supports the capacity building needed to navigate the complex ethical, environmental, and social trade-offs inherent in AI systems \cite{mulder2013sustainable, vanasupa20201c103}. Participants’ insistence that AI engineers ask how contextual experiences can be integrated speaks to the need for whole-person education approaches that cultivate both systemic understanding and deep empathy.   

\subsection{Advancing Whole-Person Education in AI Engineering}

By applying whole-person education principles to AI engineering education, this study offers several contributions to the theory and practice of whole-person education. Firstly, it demonstrates the specific relevance and urgency of whole-person education in a field fraught with complex ethical dilemmas and profound societal impacts. Unlike traditional technical curricula, whole-person education, with its focus on integrating mind, heart, and will \cite{podger2010c94d3} and fostering ethical discernment \cite{vanasupa20201c103}, provides a framework uniquely suited to address AI-specific challenges like bias amplification and accountability gaps. Secondly, this research translates foundational whole-person education concepts—such as moral motivation , honoring dignity , and systemic thinking —into concrete areas for AI curriculum development, including critical examination of neutrality, integration of design justice, context-aware ethical reasoning, and fostering empathy alongside technical skills. Thirdly, this work posits that whole-person education provides a necessary theoretical grounding for cultivating the whole engineer required for the AI era. It extends whole-person education by showing how its focus on developing higher-order dispositions , holistic well-being, and the capacity to navigate complexity is not just beneficial but essential for innovation in artificial intelligence.

%% file: 8-Conclusion.tex
\section{Limitations and Future Direction}
While this study offers rich, contextualized insights through critical auto-ethnography, we acknowledge its limitations. The sample size and specific participant backgrounds may influence the breadth of perspectives captured, and the subjective nature of the method warrants caution regarding generalizability. These limitations underscore avenues for future research. We intend to broaden participant involvement, including current AI engineering students and practitioners from more diverse geographical and cultural backgrounds, aligning with the whole-person education principle of inclusivity and listening to diverse voices (Vanasupa, 2020). Longitudinal studies are also planned to explore the long-term impact of whole-person education approaches on AI engineers' professional practices and ethical decision-making, assessing the cultivation of dispositions and potential fulfillment described by Podger et al. (2010).

%% file: PracticeDetails.tex
\section{Appendix : Extracts from Participant Reflections}

\begin{figure*}
    \centering
    {
        \includegraphics[width=\textwidth]{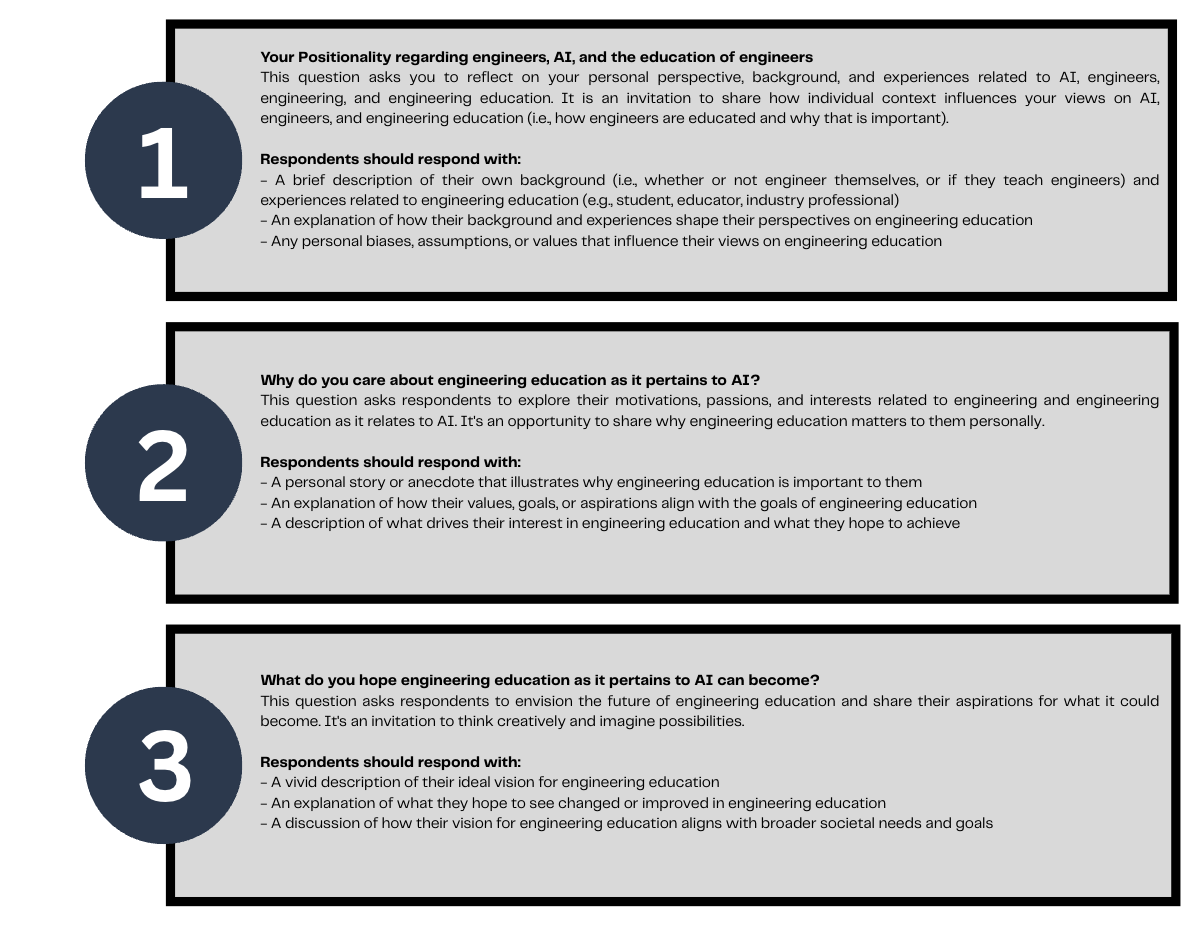}
    }
    \caption{Three prompts that each author-participant responded to - highlighting their advocacy for whole person education}
    \label{fig:enter-label}
\end{figure*}

\subsection{Rubaina Khan}
I’m Rubaina Khan and I have been in engineering education for a decade as an instructor and researcher. During this time , I have spent time designing learning experiences in the context of design and engineering ethics. In my experience, we spend a lot of time in engineering education on technical training without any context and not enough just-in-time is spent on contemplating societal implications of technology.  As a designer of experience and technology, I spend a lot of time understanding context and implications. This is often a bias because I thought many designers would do the same or my students are willing to spend that energy. This realization makes me think of ways and mandates that would make these behaviours intentional.
Accountability and responsibility with one’s professional actions should be instilled within every citizen. It shouldn’t be something that we are forced through strict regulations and mandates. Also through my educator’s experience, I noticed that mindsets toward more ethical and caring for the world behaviours needs nurturing and time. As a design educator, I see and recognise why students want to jump to creating the next new fancy tech innovation. Most of the time, it is a drag for students to think broadly and accept that certain cool features may not help or harm society or their users. It also depends on how dedicated or convinced educators are with respect to engineering for humanity. 

As I mentioned, ethical becoming needs time, this means we need to scaffold this nurturing. Teaching societal implications of technology means interdisciplinary education, dealing with uncertainty and messiness, and being aware of technical, legal, societal issues. My hope is that graduates feel they feel confident to face and recognise these challenges. Often we hear engineers feel it is out of scope or not sure how they fit into the conversation for ethical AI. Education should create this awareness and training. I hope that education in general becomes dynamic and agile that responds to all our new understanding of the world. Just as we are responding to technical education to respond to AI advancements we need to in tandem ponder and act on the implications. So that our intentions and actions are just -in-time and not an afterthought. When it comes to societal mandates, we need leaders and tech innovators to go through training - right now we just hope that they make ethical decisions. The reliance on just will to consider societal impact is alarming.

\subsection{Tammy Mackenzie}
Though I am now also a researcher in AI engineering education, I am a tech CEO by trade, and I have hired and trained about 50 engineers in our robotics lab over the last two decades. Sadly, it’s been problematic. They often don’t work well with others. Many lack basic people skills, and many technical skills. Many have been rude and resist instruction, but are also very reticent to take risks. This is across genders, cultures, sub-disciplines, ages, etc., and from multiple schools. Several of them cheated or frauded us, and we’ve even had to call the police. It has been personally very hard and mystifying to try and build small companies with engineers. We’ve lost time, money, whole teams, whole projects. I love tech. I think there’s something we’re missing in the training process so that engineers can create effectively. So that they can take on the power afforded to them by their position and do something of concrete use in a small team.  I care because society needs engineers (and everyone else) to work together, if we are to meet the challenges of our century. I hope engineering education can empower problem solvers to become good citizens. In teams, it is often hard to coordinate them with other skilled people. They tend to rely on frameworks, but be hesitant or even rude about explaining their work to their teammates. In AI, these problems are exacerbated by having to have people from many disciplines working together, as well as needing team members to think through second and third order risks. I hope that the schools can maintain the obvious curiosity and technical interests of young students without constraining them from participation in creative team work. 

\subsection{Sreyoshi Bhaduri}

As a Research Scientist at a tech company, with a background in Mechanical Engineering and a PhD in Engineering Education, I occupy a unique position at the intersection of AI, education, and engineering. My experiences as an engineer, educator, and researcher have instilled in me a deep appreciation for the complexities of developing and deploying AI systems that are both technically sophisticated, useful, future-forward, and responsible.

Through my work in engineering education, I've come to recognize the critical role that education plays in shaping the next generation of AI engineers and researchers, as well as users of engineering artefacts. I believe that AI education should prioritize not only technical skills but also critical thinking, empathy, and social responsibility. By doing so, we can empower developers and users alike to design, develop, and use systems that make lives better. As someone who has navigated the boundaries between engineering, education, and industry, I'm acutely aware of the need for more effective collaboration and knowledge-sharing across domains. I'm committed to leveraging my positionality to foster greater dialogue and cooperation between AI researchers, educators, users, and practitioners, with the ultimate goal of creating more human-centered AI solutions.

I care deeply about this issue because I believe that AI products should be designed to benefit society - this can mean different things for different people. Often, siloed, traditional evaluation metrics for AI systems can overlook the nuances of the end user experience, leading to products that may be engineering or technically sophisticated but fail to cater to user needs or sensibilities. My hope is that AI products will become more representative of perspectives, cultures, and experiences - for this, both users and developers need holistic education that empowers them and equips them with a problem solving aptitude and attitude allowing them to innovate. Finally, my hope is that as engineers and users we will all be more cognizant of our responsibilities to democratize AI solutions, and not build in siloes.

\subsection{Animesh Paul}
As a queer, international, first-generation Ph.D. Candidate in Engineering Education with a background in Electrical and Electronics Engineering, my relationship with engineering and AI has been shaped by both marginalization and possibility. I was trained in a traditional Indian engineering curriculum that emphasized rote learning and technical precision but rarely interrogated the social, ethical, or emotional dimensions of the discipline. My journey into engineering education, and more recently, into critical perspectives on AI, emerged from my desire to understand who gets to be seen as an engineer, how engineering knowledge is constructed, and why some forms of knowledge are privileged over others. Having worked in industry as a program manager and later transitioning to academia, I’ve experienced the disconnect between what engineering curricula teach and what engineering practice demands—especially in AI-intensive environments that require systems thinking, ethics, adaptability, and cross-disciplinary fluency. 

These experiences have led me to challenge the myth of the “engineer”—the idea that one can focus solely on technical mastery without engaging with the broader cultural, ethical, and social dimensions of their work. I believe engineering education must move beyond technical instruction to also include critical reflection on how technologies interact with society, whose values they serve, and whose voices are left out in the process. In the age of AI, I no longer believe that engineering education can—or should—remain apolitical. My research now focuses on equity and inclusion, with particular attention to the school-to-work transition for queer students in engineering and technologies used in engineering classrooms. I approach this work through an epistemological lens grounded in asset-based frameworks, critical pedagogy, and humanistic perspectives on technology—driven by the belief that who we are as individuals profoundly shapes what we design, how we build, and why it matters. For me, engineering must serve a more inclusive vision of society—one where technology is created for all, accessible to all, and reflective of the diverse communities it impacts, without reinforcing divides.

I care because I’ve experienced the gap firsthand. I remember being asked in a meeting to explain an analytics pipeline I had built—but no one questioned the ethics of the data used to train the model. I’ve seen both educators and students engage with AI tools and outputs without being prompted to consider the biases embedded in the underlying data. These moments have stayed with me because they reveal a critical blind spot in our educational systems: we are preparing engineers to be technically proficient, but not necessarily to be critical thinkers who question the broader implications of the technologies they create.
My motivation stems from a belief that engineering education must prepare students not just to use AI, but to interrogate it. As someone who navigates multiple identities across culture, gender, and class, I’ve seen firsthand how technologies—particularly those powered by AI—can reproduce existing inequities if we don’t teach students to ask deeper questions. I want to co-create spaces in engineering education where students are encouraged to bring their full selves, to question the values baked into their code, and to design with care and context. This work feels personal because, like many queer students I now mentor, I wasn’t always sure engineering had a place for me. Now I want to make sure it has a place for all of us.

I envision an engineering education landscape that is radically inclusive, ethically grounded, and creatively expansive—where AI is not just a tool to optimize systems, but a context to reimagine what counts as knowledge, value, and justice in technical spaces. I imagine classrooms where students are taught not just algorithm design, but algorithm accountability; not just performance metrics, but the politics of prediction. In this ideal future, engineering education embraces AI as a powerful catalyst—not just for technological innovation, but for social transformation. Students would be trained to ask: Who benefits from this model? Who might be harmed? What assumptions underlie this dataset? AI coursework would be co-taught by ethicists, community leaders, and engineers. Curricula would reflect the complexity of the real world—messy, diverse, and full of trade-offs—not just the idealized conditions of the lab. Such a vision aligns with broader societal goals of justice, sustainability, and equity. As AI increasingly shapes how we live, work, and relate to one another, we need engineers who are not just technically brilliant but morally imaginative. My hope is that engineering education evolves into a space that cultivates not just knowledge, but wisdom.

\subsection{Branislav Radeljić}

I am Branislav Radeljić, a scholar in political science and international relations. Through ongoing conversations with friends and family members working in various branches of engineering, I have come to recognize the significant value social sciences can offer to the field of engineering. While engineering plays a crucial role in addressing global challenges—whether through sustainable energy, infrastructure, or technological innovation—it is equally important to consider the broader societal implications of such progress. For instance, lithium mining is essential for advancing green technologies, yet it also raises serious environmental and ethical concerns. These issues cannot be resolved by technical expertise alone; they require a deeper understanding of social, political, and ethical contexts.This is where the integration of social sciences becomes vital. With the growing influence of AI, there is now an opportunity to embed these perspectives more deeply into engineering processes. AI can help bridge the gap between disciplines, offering engineers greater access to insights on equity, sustainability, and human dignity. Thus, for engineering to fully serve humanity, it must work in tandem with the social sciences—ensuring that progress is not only innovative, but also responsible and inclusive.

\subsection{Joshua Owusu Ansah}
I am Joshua Owusu Ansah, with a background in Computer Science and Information Technology, now pursuing a PhD in Engineering Education Systems and Design. As an African, I bring a unique perspective on inclusivity and equity in engineering education. My experiences navigating Western-centric educational systems shape my belief that curricula should reflect diverse cultural and racial backgrounds. I value diversity and believe that engineering education must evolve to serve all students equitably. My technical background also influences my systematic approach to solving these challenges. 

My mother passed away due to an inaccurate oximeter reading. I later  read a paper that talked about lack of diversity in medical device testing. This tragedy made me realize how engineering failures can disproportionately affect minoritized groups. I care about engineering education because I want future engineers to be trained to consider inclusivity in design. My goal is to help create an education system that prioritizes equity, ensuring that technologies serve all populations fairly. This personal experience fuels my passion for systemic change in engineering education.

I envision an engineering education system that is inclusive, equitable, and culturally diverse. Curricula should integrate global perspectives and move beyond U.S.-centric frameworks. Faculty should be trained in cultural sensitivity, and learning environments should foster belonging for all students. I also hope for an engineering education system that prepares engineers to design solutions that benefit diverse communities. Engineering should be a force for social good, addressing real-world challenges with inclusivity at its core.

\subsection{Indrani Bhaduri}
Through my experiences as an educator leading national, large scale education assessments in India, I have come to deeply appreciate the immense, often untapped, potential that exists within every child. In every classroom and community, I have encountered children with remarkable intellectual abilities, creativity, and a strong drive to learn—particularly in STEM areas. I have seen how engineering is appreciated and used by young learners in every corner of my country. However, systemic barriers such as socio-cultural constraints, global inequities in technology, basic lack of encouragement, increased and disproportional early responsibilities at home, and inadequate school infrastructure often stifle this potential. Despite their capabilities, many students are unable to access the opportunities or support systems they need to thrive. It has become clear to me that talent is universal, but opportunity is not, and unless we intentionally create enabling ecosystems for these young minds, we risk losing an entire generation of innovators, thinkers, and leaders.

In my extensive interactions with gifted and talented students, I have come to deeply appreciate the vast, often under-recognized and under-emphasized potential of the girl child, especially in the context of future-facing domains like artificial intelligence (AI). Despite clear aptitude and curiosity, many of these girls—particularly from underserved and rural backgrounds—face systemic barriers such as limited access to quality STEM education, societal expectations, and infrastructural constraints. In addressing these gaps, the National Achievement Survey (NAS) \cite{van2024framework}. A framework for comparing large-scale survey assessments: contrasting India’s NAS, United States’ NAEP, and OECD’s PISA. In Frontiers in Education (Vol. 9, p. 1422030). Frontiers Media SA.] serves as a crucial tool by capturing differentiated learning outcomes across demographic groups, thereby informing targeted interventions. These insights can be strategically used to identify and support girls with high potential in computational thinking, mathematics, and problem-solving—core competencies essential for future AI engineers. Moreover, the development of AI-based assessment tools that incorporate contextual variables such as socio-economic background, school resources, and cultural dynamics can further enhance the identification and nurturing of talent. Such tools not only provide a nuanced understanding of student capabilities but also create pathways for more equitable participation in AI education and careers, ensuring that talented girls are empowered to emerge as the next generation of AI engineers.

I envision engineering education evolving into a globally inclusive, age-agnostic, and gender-equitable space that not only welcomes but actively supports and uplifts every learner at every stage of their academic journey. To achieve this, we must reimagine both the content and the tools of engineering education to reflect the lived realities of all learners. This includes intentionally designing curricula, as found in research conducted while developing state-specific Holistic Progress Cards (HPCs) at the foundational [cite:  , preparatory, and secondary level. For example highlighting the contributions of global women engineers, using problem statements rooted in real-world, community-based contexts that resonate with local learners, and creating hands-on experiences that are collaborative rather than competitive. For instance, using gender-neutral robotics kits, incorporating assistive design projects that address challenges faced by women in rural areas, or using AI-augmented simulations that provide safe, exploratory learning environments can go a long way in making engineering more relatable and engaging. In addition, we also need to invest in teacher training, allowing educators time and resources to familiarize with the rapidly evolving technologies they are expected to teach engineers, fostering inclusive pedagogies and unconscious bias awareness. Only then can engineering become not just a discipline of innovation, but a vehicle of empowerment for entire communities.

\subsection{Beyza Nur Guler}
I am Beyza Nur Guler, a 3rd year PhD student in engineering education. My bachelor's degree is in civil engineering with a structural engineering specialization and my masters degree is in structural engineering. I was exposed to both the education system in Turkey and the United States and I come from a Turkish family as the oldest sibling. I also have a special needs sister and started to navigate life up close with her during my masters. In addition my father is a professor in mechanical engineering. Seeing a highly educated father and seeing him work on projects growing up pushed me to pursue higher education. In addition, experiencing both education systems, witnessing people’s financial struggles after the coup that happened in Turkey along with my struggles trying to pursue college in the United States shaped my thinking in the way that engineering education should prepare engineers to do work in industry. Moreover, during my humble efforts of trying to help my sister get an education and job in the United States I saw the barriers in educational and workplace settings for special populations. This also made me have the perspective towards engineering education that barriers should be reduced in terms of entering the workforce for anyone, in addition to retaining them in the workplace. Lastly, my first internship experience reinforced the belief that engineering curriculum should align with industry needs and the transition should be as smooth as possible. In addition, due to the type of engineering education research I have done before starting my PhD I see engineering education as an interdisciplinary field instead of a social science field applied to engineering populations. I believe Engineering Education is a bridge between educators/people who care about education to engineering faculty/ engineers. 

I have multiple stories that I can share, however the one that sticks with me the most is during a steel design exam we were asked to draw a cross section of a connection and we have never done drawing exercises before. Being able to imagine the cross section is necessary to execute the calculations correctly and design such connections currently. I was so shocked or taken aback by this experience and asked my professor how I can become better. Through multiple conversations, this ended up being my first conference paper and me learning about ASEE for the first time. I guess this was the first time I witnessed a curriculum gap. Later when I interned, I had more questions come to my head and kept asking myself why something I continuously did in industry (such as interacting with blueprints and translating them to engineering equations) was not emphasized during my undergraduate education. Even though my research is not in this space currently, I believe these memories are some reasons why I believe workforce development research is important (combined with my personal experiences of trying to get my special needs family member hired). 

Right now, AI is rapidly being adopted by businesses and it is posing barriers to people with disabilities getting hired. I believe AI skills required in industry need to be aligned with the engineering curriculum and that engineering majors that design algorithms need to also be educated about the potential risks AI can impose or exacerbate. I hope engineering education can be a bridge between disciplines, and that people who don’t care about engineering education or who have never thought of it can come to engineering education events, webinars, and conferences. I hope engineering education reaches audiences wider than the engineering education community. I hope engineering education can prepare engineers for industry and also teach them when to leverage AI, how to leverage AI and how much to leverage AI. I hope engineering education teaches how to interact with technology, how to understand what AI outputs critically while also equipping people with basic coding skills. 

\subsection{Nils Murrugarra}

My name is Nils Murrugarra. Currently, I am a Teaching Assistant Professor at University of Pittsburgh. I am teaching Computer Vision and Discrete Structures for Computer Science. I strongly believe that we have to infuse students with enthusiasm, critical thinking, problem solving skills, self-learning skills and soft skills. Topics and technologies rapidly change, and our students should develop analytical skills to understand and adapt to new discoveries in science and technology. Also, learning from others and developing effective communications skills are crucial to develop more ambitious projects, train and inspire their technological peers in the workforce.

I care about my teaching and my students. I was fortunate to have inspiring professors that even without a master or PhD, they awakened a sense of curiosity and learning to understand complex concepts. They also encourage me to pursue a research career in Brazil and the USA. I would like to give back that favor, and inspire the new generations. I strongly believe that education can unlock new opportunities and remove financial barriers from students. I was fortunate to study my master's and PhD with scholarships, and work at prestigious companies such as Snap Inc. I hope that my students have similar opportunities and they can pursue their dreams. My task is to equip them with hard and soft skills to be well prepared for the workforce.

I believe that Computer Science Education should combine technical and professional skills. From hard-skills, I value problem-solving, critical thinking, and self-learning skills. And from soft skills, we should infuse students with effective communication, team work, and social and environment awareness. In addition to technical skills, students should analyze how new developments will impact our society and our lives. Are we developing new technologies that will help us or will harm us in some way? Are they affecting our society or environment? In which ways? 

\subsection{Hayoung Shin}

I am Hayoung Shin, majored in sociology of education and educational policy, received my Ph.D., and am currently a professor of education in South Korea. Since 2010, I have been interested in the impact of digital technology on the educational environment, especially on educational resources. So I have been working on Creative Commons Korea as a volunteer and I have published papers, presented at conferences, and hosted seminars about Open Educational Resources and Open Access activities. How I get my perspectives on engineering education is as a feminist and social scientist. My main research topic is the socio-economic gap in the field of education, especially the gender gap. This chemistry has led me to have a perspective centered on Diversity, Inclusion, and Equality in engineering education. I am very interested in the gender gap and educational gap in engineering education, engineering, and STEM fields. So I studied the gender gap in the employment success of engineering college graduates, and the growth and job-change experiences of Korean startup novices according to gender.

The reason I became interested in engineering education related to AI is simple. It is the most important thing-not only in the field I study, but perhaps for all of our society. I think engineering education, if you add a little imagination, is like literacy education in the Gutenberg era. Just as literacy education allowed everyone to construct their own ideals and realities, engineering education will be an important force that allows people to communicate with the world like letters and realize themselves in this era when AI creates numerous changes and creates many problems and gaps. In my case, I try to learn their language directly and indirectly by interacting with engineering education, engineers, and engineers more than my peers in the same research field. I believe this is what makes me competitive and excellent. 

The ‘happy-harmonious coexistence’ of engineering education and AI that I envision is, above all, a happy engineering education for learners, which is engineering education that provides possibilities like a canvas for learners to create better learning experiences on their own. A painter needs a canvas to paint, and feels sufficiently free within the canvas. And when the painter’s ability is sufficiently great, he or she can also find a new and larger canvas on his or her own. 

Learners firstly think and practice with the principles and knowledge they have learned in engineering education. However, they will soon be able to imagine new engineering and engineering education. At this time, AI will act as a new color, brush, or inspiration for learners to imagine new engineering education and expand their knowledge and practice of engineering education in a more diverse way. My ideal of learner-centered engineering education ultimately corresponds to the current ‘human-friendly’ ethical and social consensus and application of AI.

\subsection{Lilianny Virguez}
I am a telecommunications engineer by training, with a Ph.D. in Engineering Education. My early career was rooted in the technical aspects of networks, systems, and communication technologies that are foundational to the digital infrastructure AI now depends on. That background has given me a systems-level view of technology and its interconnected impacts on people, society, and global development. Now, as an Instructional Associate Professor, I teach and mentor undergraduate engineering students. I also engage in cross-disciplinary research that explores the role of motivation and identity in learning. My position at the intersection of technical training and education has shaped my belief that engineers must not only master technical tools but also understand the human systems they are embedded in. I bring with me a bias toward systems thinking and a strong value for equity, ethics, and responsible innovation in both engineering and education.
My experience in telecommunications made it clear to me early on that technologies are never neutral; they carry assumptions, priorities, and consequences. That realization deepened when I transitioned into education and began helping students make sense of their roles as future engineers. AI is a powerful, fast-evolving domain, and I believe we have a responsibility to prepare engineers to critically engage with its impact, not just to use it, but to shape it wisely.
One memory stands out: a student, reflecting on their project, admitted they hadn’t considered who might be excluded or disadvantaged by the tool they were designing. That moment stayed with me. It reminded me why this work matters, to help students build not just smart solutions, but just ones. What drives me is the opportunity to help students see their work in context. I want them to realize they’re not just solving problems, they’re building futures. And with AI, that responsibility becomes even more urgent.
I hope engineering education can evolve into a space where AI is taught not only as a technical tool but as a deeply societal one where students learn how to ask critical questions, reflect on the implications of their decisions, and collaborate with people outside their discipline. I want AI education to go beyond code and computation—to include care, accountability, and purpose. I envision learning environments where students are empowered to lead with curiosity and conscience, and where diverse perspectives are central to how we define engineering solutions. I hope engineering education can support students in becoming not just users of AI, but thoughtful shapers of it.

\subsection{Rosa Paccotacya-Yanque}
I am Rosa Paccotacya-Yanque, and I hold a MSc in Computer Science. I currently teach in the Department of Computer Science at Universidad Católica San Pablo in Arequipa, Peru. Completing my master’s degree in Brazil gave me the opportunity to explore interdisciplinary applications of AI, particularly in healthcare. That experience made me realize the importance of ethical awareness, social responsibility, and critical thinking;  and now I aim to share this with my students. I want my students to think beyond the technical parts, that is thinking about the social implications, how bias can be perpetuated into systems, and how technology shapes human behavior. I have an inclination toward socially impactful projects. I think the real power of tech is in how it can help with real-world problems and actually make a difference for everybody.  I value curiosity, diversity, inclusion and  the motivation that students bring when choosing their paths.

Growing up in a region where resources and opportunities are often limited has deeply shaped my academic journey and my understanding of the transformative power of education. Since I was a child, I’ve been inspired by different role models who not only taught me technical concepts but also encouraged me to persevere and pursue my aspirations. These experiences, combined with the potential of AI when used responsibly and with ethical and social awareness, have deeply influenced how I think about education. I believe  students should be encouraged not just to code, but to think critically about why they're building something and who it’s for and how it can affect others. I want my students to see AI not only for big tech companies, but as a tool they can use to address issues that matter to them and their communities.

I hope education becomes a space that combines technical foundations,  soft skills, and a deep understanding of the ethical, social, and environmental impacts of AI systems. I hope that access to high-quality AI education becomes more equitable. I want students from underrepresented backgrounds (whether due to geography, gender, or economic limitations) to feel that they belong in this field and have the support they need to succeed. I want education to empower students not just to create innovative technologies, but to design AI systems that help to build a better world for everybody.

\subsection{Thomas Mekhaël}
I am not an engineer, but I have dedicated much of my academic and professional career to teaching future engineers. I currently hold a teaching position at an engineering school, where I lead courses on ethics, technology, and society. My work examines how engineers are formed—not just technically, but ethically, socially, and politically. In my book L’éthique et le génie québécois (PUQ, 2024), I explore the normative uncertainties and systemic blind spots that shape the professional practice of engineers in Quebec, particularly when it comes to environmental responsibility and integrity. These experiences have taught me that engineering education too often treats ethics as an afterthought—an “add-on” to technical training rather than an integral part of how engineers should think, design, and act. My positionality is shaped by both the proximity to and distance from engineering culture. I value the rigor and creativity of engineering, but I am also keenly aware of the ethical voids that can emerge when we confuse competence with responsibility. I bring with me assumptions that ethics must be more than compliance, that the social contract of engineering needs critical reflection, and that professional identity formation in engineering must be a site for nurturing democratic and ecological sensibilities—not just problem-solving skills.

What drives my concern about engineering education in the age of AI is the growing disconnect between technological capability and moral accountability. I’ve witnessed firsthand how students are fascinated by AI’s potential, but ill-equipped to reflect on its consequences—who is helped, who is left behind, and what assumptions are built into their tools. My students often want to "do good" but struggle to articulate what that means in technical practice. Their discomfort is not a failure—it’s a sign that we have work to do as educators. A moment that crystallized this for me was when I asked my class whether AI-generated solutions should be judged only by their effectiveness. The debate that followed revealed not just uncertainty, but a hunger to engage with complexity. That moment taught me that my job is not to deliver ethical answers, but to create the conditions for students to ask ethical questions. I care about this work because the power and opacity of AI systems make transparency and integrity more critical than ever. As AI takes on roles of decision-making, evaluation, and prediction, engineers must be trained to lift the hood—to make the invisible visible, and to ensure that what is being built can be understood, interrogated, and justified. AI is reshaping the very terrain of engineering practice. If we don’t anchor AI education in critical thought and collective responsibility, we risk training a generation of engineers who will build systems they do not fully understand and cannot fully justify.

My vision for engineering education is one where AI is not only a domain of expertise, but a terrain for ethical, social, and political learning. I imagine classrooms where students are trained to trace the origins of datasets, question the incentives behind optimization, and engage with voices outside their discipline—including the communities impacted by their designs. I hope we move beyond the tired binary of “hard” versus “soft” skills, and instead teach future engineers to think with both precision and humility. Engineering education should cultivate the capacity to act responsibly in situations of uncertainty, complexity, and moral ambiguity. In such a vision, transparency becomes a professional obligation, not a rhetorical accessory. We must equip engineers to document, explain, and communicate—not only their successes, but their uncertainties, limitations, and dilemmas. Only through such openness can we sustain the public trust essential to democratic engineering practice. In a world increasingly shaped by automated decision-making, AI cannot remain the exclusive domain of technologists. It must be approached as a public matter, requiring engineers who are not just technically brilliant, but socially attuned, historically aware, and ethically grounded. That is what whole-person education, to me, is ultimately about.

\subsection{Natalie Perez}
I was born and raised on the island of O‘ahu in Hawai‘i. I was formally educated in Hawai‘i, completing my bachelor, master, and doctorate degree at the University of Hawai‘i. I have an interdisciplinary educational background that includes humanities, english literature, history, and learning design and technology. My Ph.D. is in Learning Design and Technology, and I currently work as a senior research scientist for a multinational tech corporation. I do not consider myself an engineer, but I work closely with engineers, and leverage tools and technology that is common within the engineering field. My experiences with engineering education are related to my industry experience in working closely with engineering teams. Secondly, I previously supported engineering students with their writing process in generating their final capstone bachelorette projects, during my work at the University of Hawai‘i. These mixture of experiences provided me with an understanding of the type of information expected and delivered to engineering students. 

I have several personal biases and values that I acknowledge shape my beliefs and views of engineering education. Firstly, having been trained in both the humanities, education, and technology spaces, I value interdisciplinary and cross-discipline education and learning experiences, and I also see the friction between the humanities and technology spaces. At times, I feel this friction socially is defined as art vs. science, rather than my viewing of the fields as being both art and science. Along these lines, my educational upbringing has instilled in me the importance of philosophical questioning and considering ramifications or potential ramifications of any actions or decisions, regardless of field. My educational experiences have also positioned me to question the function, use, and purpose of technology in particular, including its potential impact on learning, culture, communities, and society at large. In the Hawaiian islands, there is a consistent push to question how technology is supporting indigenous peoples, such as the Kanaka Maoli people and Hawaiian culture. One example are efforts to improve AI through indigenous knowledge. Consequently, I view engineering education through an interdisciplinary lens that equips engineers to not only have the knowledge, skills, and abilities to build, but also to think critically, examine implications, and create through a lens of ethics, transparency, and human-centered designs that consider communities, culture, and society. 

When I consider engineering education and what is important to myself with respect to AI, I reflect on ethical concerns that include privacy, transparency, truth-telling, security, and broader cultural and societal concerns like representation, technical barriers, and job availability. More broadly, I think about what it means to be human. The infamous statement by French philosopher, Rene Descartes, said, “I think, therefore I am.” This statement reinforces that the act of thinking is proof of the reality of one’s mind. Yet, if we consider the context of AI, and its semblance of “thinking,” I reflect on its potential to be a Faustian Bargain. For instance, as Faust gives up his mortality to Mephistopheles, he releases AI, surrendering his control as a human to a machine. In light of these reflections, it is clear that engineering education must not only prepare students to advance technology but also to critically evaluate its impact on humanity. As AI continues to blur the line between human and machine “thinking,” the role of engineers becomes even more vital—not just as builders of systems, but as stewards of ethical and cultural responsibility. By grounding engineering education in human-centered values like transparency, representation, multiculturalism, and truth-telling, we can work to the essence of what it means to be human: to think, to observe, to control vs. be controlled. To avoid a Faustian Bargain—where we exchange our agency for convenience—we must equip future engineers with the tools, thinking, and ethics to maintain both control and agility in a rapidly evolving world.

My ideal vision for engineering education in the age of AI is one where students are empowered not merely as users of advanced tools, but as thoughtful co-creators in shaping a shared future with intelligent systems. I imagine classrooms that go beyond coding and algorithms—spaces where philosophy, ethics, cultural studies, and human behavior are woven seamlessly into the curriculum. Here, students learn not just how AI works, but what it means to build alongside it. We should challenge students to not just ask: Can I build this? But, more importantly: Should we build this? 

\subsection{Curwyn Mapaling}
I am not an engineer. I am registered as a clinical psychologist, trained in the language of psyche, emotion, and meaning-making. My roots lie in the social sciences, and yet, for several years, my academic life has unfolded in the corridors of engineering education. At Nelson Mandela University, I co-developed and taught a Critical Thinking module for first-year students (engineers-in-the-making) and later facilitated a Business Engineering course at the second-year level. For two years, I served as an academic advisor: first to the School of Engineering, and later to the entire Faculty of Engineering, the Built Environment and Information Technology.

Being situated as an “outsider” in engineering spaces has allowed me to witness the edges, what is centred and what is often overlooked. My training teaches me to attune to the unspoken, the affective, the deeply human. I’ve seen how this can sometimes be missing in the rigour and precision of engineering curricula. But I’ve also encountered engineers whose capacity for empathy outpaces that of some in my own discipline, reminding me that emotional intelligence doesn’t live in any one profession.

My relationship with engineering is not only professional; it is also deeply personal, in actual fact, it is familial. My father dreamed of becoming an electrical engineer. That dream never fully materialised, but he spent much of his career teaching vocational subjects to engineering students at a Technical Vocational Education and Training (TVET) college before moving into management. He also holds an honours degree in educational psychology. I didn’t inherit his mathematical gift, but I suspect I carry his elements of his unfinished dreams. My openness to working with engineers, inside and outside the classroom, is, in part, an extension of his legacy. In a way, I’m continuing his work through a different portal.

My perspective on engineering education, then, is shaped by this braid of histories, my own disciplinary training, the relationships I've built in engineering spaces, and my father's quiet influence. I bring to this work a bias toward the human. I am suspicious of logics that erase context. I value emotional depth, cultural sensitivity, and the messiness of being. These are not always the loudest values in STEM education, but I believe they matter, perhaps now more than ever, as we educate engineers to face an increasingly complex, interconnected, and fragile world.

It was during my time as an academic advisor to the Faculty of Engineering that something quietly took root. I was not meant to stay long, psychologists don’t typically linger in engineering corridors, but the students kept pulling me in. Their questions, their quiet fatigue, the tension behind their resilience. I began to notice a particular kind of grit in them. Not loud or boastful, but present. Persistent. Despite full timetables, dense scientific material, and institutional cultures that were often unforgiving, they were showing up. Pushing through. Bouncing back. I wanted to understand this, what it meant to be academically resilient within high-risk disciplines, and what it costs.

That wondering became the seed of my PhD: an exploration of the academic resilience of engineering students. What began as a curiosity about how they cope turned into a deeper inquiry into what they carry, how they’re supported (or not) and how they find ways to remain afloat. It was never just about survival for me, but about the human within the engineer. Over time, that journey evolved into work with engineering educators, probing the discourses they use about teaching and learning, and how these, in turn, impact their own well-being. I began to see how tightly wound the system is, and how essential it is to make space for breath, for care, for rehumanising the educational experience.

I am drawn to engineering education not because I want to make engineers out of everyone, but because I believe that who becomes an engineer matters. What kind of engineers are we cultivating? What values shape their designs, their decisions, their sense of accountability to the world? I want to be part of a generation of educators and researchers committed to cultivating socially responsive engineers, engineers who will wield AI with ethical intentionality, who understand the weight of their calculations, and who refuse to forget the moral psychology of their craft.

What drives me is the hope that engineering education can be more than just technical proficiency. That it can nurture critical, compassionate thinkers who recognise their work not as separate from the social world, but embedded within it. I hope to contribute to a vision of engineering education that is rigorous, yes - but also just, soulful, and alive to the complexity of what it means to build, to design, to engineer in a world that needs more healing than harm.

I imagine an engineering education that no longer positions the technical and the human as binaries. Where AI is not only a tool to be mastered but a mirror held up to society, to ethics, to ourselves. I dream of classrooms where code is taught alongside conscience, where algorithms are dissected with as much care as affect, and where engineers are asked not just what they can build, but why, for whom, and at what cost.

In this reimagined future, engineering education moves beyond efficiency and productivity and learns to linger. To hold space for grief, for wonder, for difficult conversations about race, gender, inequality, and environmental collapse. I hope to see a curriculum that breathes, that welcomes the arts and humanities into its chambers not as visitors, but as core occupants. I want to see engineering students reading bell hooks, Toni Morrison and Maya Angelou alongside technical manuals, sitting with the poetry of physics and the politics of design.

AI, in this vision, is taught not only as innovation but as inheritance, built on the backs of many, shaped by the choices of few. I want engineering students to learn about the biases that travel through datasets, the silences coded into seemingly neutral technologies, and the lives altered by “intelligent” systems. And more than that, I want them to care. I hope we begin to prioritise emotional literacy alongside computational literacy. That we resist the temptation to make future engineers more machine-like in their precision and instead invite them to be more human, curious, fallible, responsible, and imaginative. I want them to know that building bridges and designing systems is also about building trust, designing for justice, and staying tender in a world that hardens too quickly.
What I long for is not merely a transformation in content, but in consciousness. An education that does not evacuate the soul from science. One that aligns itself with the broader project of social justice, ecological restoration, and human dignity. An education that does not treat AI as destiny but as possibility, one to be shaped with humility, solidarity, and care. That is the engineering education I hope for. One where the question is not only what can we do with AI, but who do we become and unbecome because of it?

\subsection{Debarati Basu}

As an international woman faculty member from a country rich in cultural, religious, and racial diversity, my academic journey in engineering education has been shaped by a unique blend of cultural and disciplinary perspectives. I hold undergraduate and graduate degrees in Computer Science and Engineering, and a Ph.D. in Engineering Education. I gained industry experience as an Assistant Systems Engineer focused on web development, along with academic experience through teaching and research during my Ph.D. For the past seven years, I have served as a faculty member with a research focus at the intersection of engineering and computer science education, teaching courses ranging from introductory to graduate levels.
These diverse experiences have significantly shaped my views on engineering education and its relationship with AI. First, I believe it is essential for students in computing and engineering to understand the responsible use of AI, especially in academic contexts, which is reflected in my work on developing and evaluating module on academic integrity for computing students. Second, I strongly advocate for the inclusion of diverse perspectives in AI technology development to reduce algorithmic bias and promote equitable outcomes. Third, in teaching, I see tremendous potential for AI to support instructional design and content generation, provided it is used ethically and thoughtfully. Finally, in education research, AI tools can be transformative—but only if they are developed to be inclusive and unbiased, ensuring they serve educators and learners from all backgrounds. Overall, my values emphasize equity, responsible innovation, and the transformative power of education, principles that continually inform my approach to AI and engineering education.

My interest in engineering education and AI stems from both my professional journey and personal experiences. In my classrooms, I often encountered students who used AI tools to complete a coding assignment without grasping the core concepts. These moments make me realize the urgent need to teach students not just how to use AI, but how to do so responsibly and thoughtfully. With a background in both industry and academia, I see engineering education as a space to develop not only technical skills but also ethical awareness, and critical thinking. My values in equity, integrity, and innovation, drive me to create learning environments where students critically engage with AI, understanding its potential and its limitations, and use it responsibly. I care deeply about helping students become not just capable engineers, but also responsible ones. Ultimately, I care about engineering education because I believe it can empower diverse learners and foster responsible engineers who use AI to create inclusive and meaningful impact.I envision an engineering education landscape where AI is not just a tool for efficiency but a catalyst for deeper learning, equity, and innovation. In this vision, AI helps tailor education to individual learners, supports instructors in designing adaptive and inclusive content, and empowers students to become both creators and critical consumers of technology. 

I hope to see a shift from content-heavy, one-size-fits-all instruction to more dynamic, AI-assisted learning environments that promote critical thinking, creativity, and ethical decision-making. Engineering education should prepare students not only to build AI technologies, but also to question their implications and ensure they serve the public good. This vision aligns with broader societal needs, equity in education, responsible innovation, and preparing a diverse generation of engineers who can lead in a rapidly evolving technological landscape. I believe AI, when thoughtfully integrated, can help transform engineering education into a more inclusive, responsive, and socially conscious system.

\subsection{Rodney Kimbangu}
I hold an associate degree in civil engineering, a bachelor's degree in painting and film, and a Master of Fine Arts in Creative Technologies. From my perspective, the latter integrates both modes of thinking, engineering, and artistry. Interestingly, I now work as a communications practitioner at the Institute for Creativity, Arts, and Technology (ICAT), where I help communicate across the intersections of art, design, and technology. This unique convergence gives me insight into systems' structural logic and creative technologies' expressive, human-centered dimensions. It has shaped my belief that neither engineering nor art should operate in isolation because they are deeply complementary ways of knowing and shaping the world. The only problem is that in my time being trained as an engineer, ethics were glossed over, while in the arts, it was embedded in the language used and even most of the practice itself, even though sometimes it was twisted to fit political agendas. I also recognize that my artistic training may lead me to emphasize humanistic and interpretive aspects of engineering more than someone with a purely technical background might. I tend to be skeptical of engineering approaches that prioritize efficiency and innovation without deeper ethical scrutiny or emotional resonance.

Engineering is a powerful discipline that allows us to reshape the world around us. Much of our economic and technological progress can be directly or indirectly attributed to engineering practices. With the advent of artificial intelligence, this power has been exponentially magnified. AI opens Pandora's box of ethical dilemmas and holds immense potential to support a more equitable and sustainable future, but only if we choose that path. My background in engineering taught me to approach problems with logic and efficiency, but it often overlooked the more profound implications of design decisions, humans. I still recall a project during my early civil engineering studies where I was proposing to build a road for an underserved part of the city, and all I cared about was imagining and building the road, how much it would cost, and how I could have optimized the structural design without once discussing who would use it, or what communities it might serve. It didn't seem odd at the time, but looking back, I see that this technocentric training left little room for ethical imagination.

This absence is even more dangerous in AI development, where systems learn from historical data and make predictions that directly affect people's lives. It's no longer enough to just build efficient systems; there is a serious call to question what those systems represent and whom they truly serve.

I hope that engineering education will initiate conversations that many engineers either don't realize are necessary or have been taught to ignore about ethics, user impact, and broader societal consequences. With powerful and rapidly evolving technologies like AI, we need to shift engineering mindsets beyond function and performance toward reflection: Who is affected? What values are encoded? What long-term ramifications might emerge?
Too often, engineers are expected to produce "the thing," a functional system, a product, a tool, all without interrogating its deeper implications. However, that expectation aligns more with how we traditionally treat art, where meaning is left open to interpretation. Ironically, artists are frequently asked to think beyond their work to justify, critique, and contextualize it. Why shouldn't we ask the same of engineers whose creations are embedded in everyday life and carry real-world consequences? We design and engineer technologies for people. Therefore, engineers must be invited and required to think about what they are building, who will use it, how it will be used, and who might be left out or harmed.

This is why I align with the call for a whole education approach in AI engineering, cultivating ethical stewardship and supporting interdisciplinary learning as core components rather than peripheral concerns. Such an approach reminds us that engineering is never value-neutral; it is always entangled with questions of justice, access, and human well-being. I envision engineering classrooms where students work on real-world challenges in diverse teams, engaging with community partners and exploring not just how to build systems, but whether they should be built them, and for whom. Ethical reasoning, communication, and reflective practice would be as central as algorithmic performance, and would embedded across courses, not treated as electives or afterthoughts. In an AI-driven world, engineering education must become as much about people as it is about precision.

\subsection{Leslie Salgado}
I’m an outsider to the engineering field. However, as a science communicator and scholar in the Studies of Science and Technology Field,  I have been following for a long time various issues pertaining the engineer field:  how creative thinking fuels engineering process, how design choices entail and reflect political decisions, and how participatory options from ideation process can contribute to democratize the technical code. I’m a humanist who strongly believes in the power of embedding justice in design, and enjoy the interdisciplinary conversation.

While AI-powering technologies are being progressively embedded into everyday practices, they are becoming infrastructure. The way AI is transforming processes and everyday artifacts design should be a question at the core of the engineering classroom. The change in the human-artifacts relationship triggered by AI-powered technologies cannot be overlooked by engineers educators. Bringing those debates to the classroom is essential to demystify technological neutrality and set the basis for design practices capable of supporting the pathway towards social justice. Although this has been explicitly tackled for decades in engineering education, AI makes that need more evident than ever. The scale, scope and pace of the transformation fueled by AI is what makes this an urgent need. 

First, I’m hoping that engineering education can provide space to nurture non-orthodox approaches to creativity. In the midst of the “AI revolution”, I’m hoping that future engineers are educated in creative and critical thinking. With the current state of things– AI worsening inequality– is not an option to encourage a thinking that interrogates the realities before the design process. It is imperative to explicitly teach how AI is impacting social injustice and inequalities. This entails an approach where ethics, and social studies are not relegated to one course-two courses, but span across the whole curriculum. 

To promote creative thinking that goes beyond traditional design practices, going back to the classics, not as a History of engineering course, but as a philosophical stance, would represent a good step forward. Examples from the Talos automaton to Ismail al-Jazari could help students to understand how arts, creativity, and intellectual reflection have been at the core of engineering. The key, to me, is to promote the potential of alternative technological pathways.

\subsection{Yves Gaudet} 
I am the owner and principal operator of a factory. We recycle plastic and mass produce molded plastic items.  I’ve been in this business, and making our own robots since I was a teenager. Now almost 40 years. The fact is: I have trouble working with engineers. They don’t integrate well on the floor with the operators who know the machines, and they often will jump ahead without knowing or asking, sometimes assuming this to be a professional responsibility, even when dealing with people who are better informed. I think they are great with details and frameworks, but in practical terms they really have trouble being flexible. As a medium enterprise, that type of flexibility and thinking of how our work affects society is really key to our business model. As we put more and more automation into our systems, I hope to see new engineers with more interpersonal and interdisciplinary skills, more confidence to take some chances, and better technical skills to match their theoretical tool boxes.

\subsection{Allen Munoriyarwa}
I have established myself as a researcher, some of whose work is at the nexus of AI and media. This means I researched how AI affects specific communities of practice, such as media workers (journalists), etc. Therefore, my research always deals with the by-products of AI engineering. 

I understand how AI affects certain communities of practice, and the question of AI and the humane is at the centre of my work. How can an ideal AI integrate ethics and contextual differences? How can AI help create a humane society? 

These are important questions for AI engineers that are rarely asked, but are important for AI engineers to remember. Coming from Africa means we often have to deal with AI technologies largely imported and made from our cultural, socio-economic contexts.This raises the question of whether some African societies will be marginalised in this AI-driven society. This raises the question: Are AI engineers aware that an AI-driven society should be driven by equality and equity, as well as an acute awareness of cultural sensitivities. This drives my understanding of AI and what Engineers would do about it. The future of engineering education should put the values of equality, inclusion, and cultural sensibilities in the matrix of AI engineering. 

\subsection{Paula Larrondo}
I am a geologist with an MSc in Mining Engineering (Geostatistics) and now a PhD candidate in Engineering Education, studying how AI can provide feedback to engineering students in solving complex problems. I arrived at this point largely because of my daughter’s ADD, which I most likely inherited to her. Watching her struggle in a system built for neurotypical learners reshaped my idea of success profoundly and inspired my commitment to education that values diversity and collaboration. Years ago, I was given the rare opportunity to found and direct a school of geology at a private university in Chile. My goal was to challenge the  Chilean stigma at that time, that private institutions offered inferior training, and to open quality programs to less privileged students. Reality, however, quickly humbled my idealized view of educating others. Teaching students who may have been underserved in so many different ways by an unequal educational system requires the best teaching skills, and I am still in the process of earning those stripes.

While doing my best to assess my students’ learning properly, I noticed something striking. Students often believed their answers matched the exemplar I used for marking and that they could review as part of the feedback process, yet their reasoning contained hidden gaps. Geology, like engineering, is full of complex relationships; only when we ask, “why does X or Y happen?” do we expose those gaps and misunderstandings. Written explanations became my lens for revealing—and helping students repair—those conceptual faults.

Written communication is a window to our understanding, or the lack of it, when faced with a complex problem. And although AI, particularly Large Language Models (LLMs), cannot “understand” a problem, they are remarkably good at spotting holes and inconsistencies in our writing, and, by extension, logic. 

The real challenge is crafting prompts that elicit valuable insights and structuring the AI’s responses, so they guide students toward deeper reasoning rather than handing them an instant solution
Biases? Absolutely. I am a neurodivergent, second‑language writer with a language‑processing disorder who relies on LLMs every day. These tools have transformed my scholarly life. Plus, I’m a little obsessed with making myself as straightforward as possible. I don’t think you can be more biased than that.

My hope is that students learn to use AI as a tool to become better at complex reasoning and communicating, not at doing the work for them. However, this is something the students need to acquire through guidance. Letting them learn this by themselves might not be the most effective or efficient way of doing it. Like teamwork skills, effective AI use demands explicit instruction, not mere evaluation. I also believe ethical deployment of these tools can level the playing field and help neurodiverse learners thrive. When diverse minds join the engineering workforce, society benefits from the broader range of ideas and solutions they bring.